\documentclass[11pt]{article}

\usepackage{float}
\usepackage{amsfonts}

\usepackage{amsmath}
\usepackage{amssymb}
\usepackage{latexsym}
\usepackage{graphicx}
\usepackage[english]{babel}
\usepackage[font={small}]{caption}

\usepackage{amsfonts}
\usepackage{latexsym}
\usepackage{graphicx}
\usepackage[english]{babel}
\usepackage{tikz}

\begin{document}
\input epsf

\def\p{\partial}
\def\h{{1\over 2}}
\def\be{\begin{equation}}
\def\bea{\begin{eqnarray}}
\def\ee{\end{equation}}
\def\eea{\end{eqnarray}}
\def\d{\partial}
\def\la{\lambda}
\def\eps{\epsilon}
\def\b{\bigskip}
\def\m{\medskip}

\newcommand{\newsection}[1]{\section{#1} \setcounter{equation}{0}}

\def\q{\quad}

\def\h{{1\over 2}}
\def\t{\tilde}
\def\r{\rightarrow}
\def\nn{\nonumber\\}

\let\p=\partial

\newcommand\blfootnote[1]{%
  \begingroup
  \renewcommand\thefootnote{}\footnote{#1}%
  \addtocounter{footnote}{-1}%
  \endgroup
}

\begin{flushright}
\end{flushright}
\vspace{20mm}
\begin{center}
{\LARGE Three puzzles in cosmology}
\\
\vspace{18mm}
{\bf    Samir D. Mathur }\\
\vspace{10mm}
Department of Physics,\\ The Ohio State University,\\ Columbus,
OH 43210, USA \vspace{2mm}\\ mathur.16@osu.edu

\vspace{8mm}
\end{center}

\vspace{4mm}

\thispagestyle{empty}
\begin{abstract}

\vspace{3mm}

Cosmology  presents us with several puzzles that are related to the fundamental structure  of quantum theory.  We discuss three such puzzles, linking them to effects that arise in black hole physics. We speculate that puzzles in cosmology may be resolved by the vecro structure of the vacuum that resolves the information paradox and the `bags of gold' problem for black holes. 

\end{abstract}
\newpage

\setcounter{page}{1}

\numberwithin{equation}{section} 

\section{Introduction}

\b

Cosmology today presents us with a fascinating intersection of experimental data, theoretical models and deep puzzles. In this article we describe some of the puzzles that arise when we try to relate the broad features of the observed universe to fundamental principles of physics.

At first one might think that cosmological physics need not be concerned with any gaps in our understanding of physics at the planck scale. After all, cosmology involves very large length scales where the standard rules of classical gravity and quantum field theory seem to work adequately. But the cosmological constant problem already tells us that it is not so easy to separate scales:  the small average curvature of the universe seems to require a fine tuned ultraviolet cutoff in the field theory.  

As we will note below, a second set of problems follow from the difficulties associated with black holes. Black holes are macroscopic objects, but understanding their entropy and radiation has been difficult. First, the standard semiclassical picture of the hole does not manifest the degeneracy of states required to account for the entropy. Second,   radiation from the hole leads to the well known information paradox. 

 The fuzzball paradigm for black holes  in string theory radically modifies the structure of the hole: the mass is not concentrated at a central singularity, but is instead spread over a horizon sized object termed a `fuzzball'. The number of these fuzzball states accounts for the entropy, and any individual fuzzball state radiates from its surface like a normal body, thus evading the information paradox. Does this change in our picture of the black hole have any consequences for the large scale structure of the universe? 

The classical dynamics of  the big bang is similar to the time reversed evolution of a star undergoing gravitational collapse. Thus any novel physics that alters macroscopic physics for black holes can also be relevant to cosmology. In fact as we will note, the information puzzle can be mapped to a corresponding puzzle in cosmology, using the Birkoff theorem. 

It is fascinating that these relations between black hole puzzles and cosmology opens up a path to using cosmological observations to constrain models for the black hole interior. Conversely, any understanding of the black hole horizon has a relation to the expansion of the universe on the cosmological horizon scale. 

\b

It is a pleasure to be able to contribute my thoughts on these issues to this volume commemorating the excellent review \cite{review} by Copeland, Sami and Tsujikawa. This review served as a starting point for my understanding of what is known about cosmological data and theory. What I seek to do here is to recall three basic puzzles about cosmology which must be resolved in a manner consistent with the observations recounted in articles like \cite{review}. 

\b

In brief, we will discuss the following issues:

\b

(A) As we look back towards the big bang, the energy density of matter becomes very large. Does fundamental physics predict an equation of state $s(\rho)$ in the limit of arbitrarily high densities? (Here $s$ is the entropy density, and $\rho$ is the energy density.)   Two very different lines of argument lead to the same result
\be
s=c\sqrt{\rho\over G},~~~c\sim 1
\label{one}
\ee
One line of argument uses black hole thermodynamics: we ask what arrangement of black holes can lead to the maximal entropy for a given energy $E$ in a given volume $V$. This leads to the `black hole gas', with the equation of state (\ref{one}). The second argument starts with string theory, and notes that there is a unique expression for the entropy in a box that is invariant under the S and T duality symmetries of the theory, and that matches the entropy of a black hole in an appropriate limit. Remarkably, we again find (\ref{one}).

We should therefore ask: does the equation of state (\ref{one}) correctly describe the dynamics of the very early universe? If not, how do we escape the very general nature of arguments that lead to  this equation of state?

\b

(B) The semiclassical picture of the black hole leads to the information paradox. In cosmology, the mass $M$ inside the cosmological horizon radius $R_c$ is of the same order as the mass which  would make a black hole with radius $R_c$ in asymptotically flat space. Should we worry then about a version of the  information paradox associated to the cosmological horizon?

Hawking radiation is a low energy quantum process. So one might think that subtle effects in the quantum theory will somehow resolve the information puzzle, and we need not worry about these effects when studying the much higher energies studied in cosmology. But the `small corrections theorem' \cite{cern} tells us that the puzzle {\it cannot} be resolved if we preserve the basic aspects of gravity at low energy scales.  We have two choices:  (i) We violate the semiclassical approximation by order unity at the horizon, or (ii) we invoke very nonlocal effects caused by entities like `wormholes' which transport information from inside to hole to distant regions. In either case we must ask: are these violations of normal local semiclassical physics observed at the cosmological scale? If not, why not?

If we assume that our full theory of quantum gravity respects causality, at least to leading order, then the information paradox gives strong constraints on the nature of the quantum gravity vacuum. It appears that the only way to satisfy these constraints is to assume that the vacuum contains fluctuations of `virtual black holes' for holes with all radii $R$.  The natural suppression of such fluctuations for objects with 
$R\gg l_p$ is offset by the large degeneracy of black hole type states for large $R$. The part of the vacuum wavefunctional describing such fluctuations is called the `vecro' component \cite{vecro}. Altering this component at length scales $R$ can give an effective cosmological constant  $\Lambda \sim (GR^2)^{-1}$, which has an energy density of order the closure density. It thus seems possible that both the energy needed for inflation and the dark energy observed today might arise from the dynamics of this vecro component of the vacuum. Thus we should ask: is there evidence for such vecro dynamics in the sky? If so, can such dynamics be consistent with the constraints provided by BBN, structure formation and BAO observations?

\b 

(C) We understand particles as excitations of a quantum field filling all of space. But the vacuum energy density of the field diverges when we add up the contributions to the vacuum energy from short wavelength modes; this is the cosmological constant problem. It appears that one should first cut off the divergence at the planck scale to get a finite energy density; one could then worry about making this energy density small.

Suppose we consider a scalar field and  implement the cutoff by placing the field on a lattice with spacing $l_p$. The energy density of the vacuum is now $T_{tt}\equiv \rho\sim l_p^{-4}$. But the stress tensor does not turn out to be proportional to $\eta_{\mu\nu}$; instead one finds pressures $T_{ii}\equiv p=\rho/3$, as expected for a massless gas. So before we can worry about making $\Lambda$ small, we are faced with the difficulty that we do not even have a $\Lambda$; we have instead broken Lorentz invariance. 

One might try to remedy this by using dimensional regularization, which by construction gives a Lorentz invariant result. But this regularization is a purely formal procedure, and suggests no direct map to the fundamental way in which the theory achieves its finiteness. 

The difficulty with the lattice cut-off is quite fundamental. We define $T_{\mu\nu}=-{2\over \sqrt{-g}}\delta S/\delta g^{\mu\nu}$. But when we vary the spatial metric $g_{ii}$, we increase the size of the spatial directions, and this increases the lattice spacing. To get a Lorentz invariant stress tensor, we should add more lattice sites when we increase $g_{ii}$, so that we maintain the lattice spacing at the value $l_p$. This feature would extend to cosmological evolution: we should add more lattice sites as the universe expands.

But now we face a problem: if we increase the number of lattice sites, then we increase the dimension of the Hilbert space. But Hamiltonian dynamics works on a Hilbert space of {\it fixed} dimension. 

Thus we should ask: Is there a formulation of quantum theory where all quantities are finite and well defined, yet the vacuum energy is proportional to $\eta_{\mu\nu}$ and the spacetime can expand? Or is this difficulty pointing to a fundamental gap in our understanding of vacuum energy? We have noted above that the information paradox indicates the existence of  a `vecro component' in the vacuum wavefunctional which describes fluctuations of virtual black holes of all sizes. Could it be that it is these large scale fluctuations that temper the divergences of the field theory, and the issue was never about implementing a planck scale cutoff?

We now discuss the puzzles (A)-(C) one by one.

\section{The equation of state at high energy density}

Consider a flat cosmology in $d+1$ spacetime dimensions
\be
ds^2=-dt^2+a^2(t) \sum_{i=1}^d  dx_i dx_i
\label{cosmology}
\ee
Let us compactify the spatial sections to a torus $T^d$; if our entropy turns out to be extensive, then we can replace the torus by uncompactified space. 

Let us now ask the following question. Suppose we have a toroidal box of volume $V$. In this box we put an energy $E$. What is the entropy 
\be
S=S(E,V)
\ee
in the limit $E\r\infty$?

Before proceeding to find $S$, let us check if we are asking a reasonable question. Thermodynamics tells us that a system, left to itself, heads towards the state with maximal $S$ for given parameters $E,V$. But the notion of inflation in cosmology has led us to think of states which have {\it low} entropy. The underlying reason for this difference  is the negative sign of the gravitational potential. This negative sign allows  metastable configurations where the universe expands quickly and increases the matter part of the energy, which is  compensated by the negative gravitational potential energy. The state at the end of inflation is a state with low entropy for its total matter energy.

But apart from the situation where we are trapped in such inflationary expansion, we {\it do} consider states for the universe  where the entropy is maximized. In early work on the big bang we considered the early universe to be filled with radiation, since radiation maximized the entropy in the standard model at high temperatures. When we learnt about strings, we  considered the entropy of a string gas at very early times \cite{bv}. When we discovered  branes, we looked at the maximal entropy state of a gas of branes carrying  vibrations on their surface \cite{branegases}. Thus  apart from  periods of inflation,  we do  look for maximal entropy  configurations in the early universe. 

Note that taking the  thermal state at an  early time $t$  does not  mean that we will not see entropy increase at  later times. In the thermal state at time $t$ we  have the maximal allowed entropy in the volume $V(t)$ that the  universe had at  time $t$; as the  universe expands, $V$ increases, and then entropy can  increase further, as  we observe  in the world around us today.

\subsection{The maximal entropy state}\label{secmaximum}

Our present aim is to consider the universe right after the big bang, before inflation, and it seems reasonable to ask what  the maximal entropy state is at these early times. Thus let us ask: what equation of state will we get when we look at the maximal entropy configurations of our quantum gravity theory in the limit  $E\r\infty$ in any given $V$?
 
\b

\begin{figure}[h]
\begin{center}
\includegraphics[scale=.72]{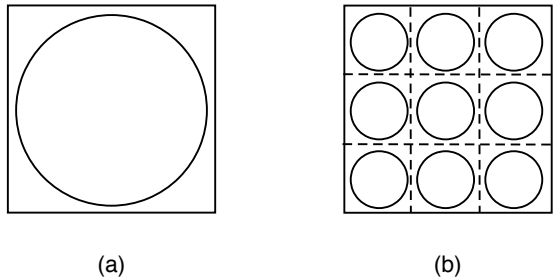}
\end{center}
\caption{\label{fbh} (a) At energy $E\sim E_{\rm bh}$ the maximum entropy is attained by a single hole filling the box (b) At larger $E$, a lattice of black holes has more entropy than a single hole.}
\end{figure}

For low values of the  $E$,   we  expect the phase of matter to be radiation. This phase has  entropy $S\sim V\rho^{d/ (d+1)}$, where $d$ is the number of spatial  dimensions.

 At larger $E$, we can get more entropy by forming a black hole. A black hole looks like a somewhat esoteric object, and one might wonder if it should be considered in our search for the phase with  maximal entropy. In classical gravity, black holes can form but not `un-form', so if we start with all the matter in black holes then we could worry about how we would ever exit such a phase. But we have have learnt  from string theory that the Bekenstein entropy \cite{bek} 
 \be
 S_{bek}={A\over 4G}
 \label{sbek}
 \ee
   is reproduced, at weak coupling, by counting the intersections of branes \cite{sv}. 
 When we increase the coupling to the point where we expect black hole formation, these microstates turn into horizon sized quantum objects called `fuzzballs'; these are solutions of the gravitational theory that are in principle just like any other object in the theory like a star or a planet \cite{fuzzballs}. Thus we should consider states containing black holes  as normal states to be counted in the search for the maximally entropic configuration.

As we increase the energy  in the box  we reach a value 
\be
E\sim E_{\rm bh}
\label{oone}
\ee
 where the radius of the hole $R$ becomes order the size $L$ of the box (fig.\ref{fbh}(a)). The entropy of this black hole is 
\be
S_{\rm bh}\sim  {A\over G}\sim {L^{d-1}\over G} 
\ee
What happens if we increase $E$ further? At first one might think that there could be some physical principle which does not allow a larger $E$ in the box, since a larger black hole will not fit in the box. But classical general relativity places no bounds on the energy in a box; a larger energy just implies a larger Hubble expansion rate
\be
\left ( {\dot a \over a}\right )^2={16\pi G \over d(d-1)}\rho
\label{otwo}
\ee
Further, suppose we take a particular time $t$  in the radiation phase of the cosmology (\ref{cosmology}) with energy density $\rho$ and entropy density $s$. In a ball of proper radius $R_b$, the energy is $E\sim (R_b)^d\rho$, corresponding to a black hole Schwarzschild radius 
\be
R\sim (GE)^{1\over d-2}\sim (G\rho)^{1\over d-2} R_b^{d\over d-2}
\ee
For large enough $R_b$, we will have $R>R_b$. So we can obviously have more energy $E$ in a box of size $R_b$ than would be allowed if we insisted that the energy in a box be bounded by the mass of the largest hole that can fit in the box. 

So we consider energies $E>E_{bh}$ and ask what entropies we can achieve in this energy domain. Let us still try to use black holes in the volume $V$, but now with this larger energy $E$. A simple construction shows how we can place energy $E>E_{bh}$ in our box. Instead of taking one black hole filling $V$, we consider a lattice of smaller holes (fig.\ref{fbh}(b)), each hole with the same radius $R$. Let the  separation between the holes also be $\sim R$. The number of holes is then
\be
N_{\rm hole}\sim \left ( {V\over R^d}\right )
\ee
The entropy of each hole is
\be
S_{\rm hole}\sim {R^{d-1}\over G}
\ee
Thus the total entropy is
\be
S\sim N_{\rm hole} S_{\rm hole}\sim {V\over R G}
\label{entropylattice}
\ee
We see that we can make $S$ as big as we want by making $R$ small enough. In particular,  the entropy of such configurations will in general exceed the entropy given by the surface area of the box. Imagine each hole of radius $R$ to be surrounded by a cubical box with sides $2R$; the horizon area $A$ of the hole in each box is then, upto a factor of order unity, the area of its cubical enclosure. We see that the outer faces of the outermost boxes yield an area that is of order the horizon area of the hole in fig.\ref{fbh}(a). The other surfaces of the cubical boxes in fig.\ref{fbh}(b) contribute yet more area, so we see that the total entropy in fig.\ref{fbh}(b) can be made much more than the entropy in fig.\ref{fbh}(a).\footnote{I first learnt of this simple argument from Ali Masoumi.}

 The energy of each hole is
\be
E_{\rm hole}\sim {R^{d-2}\over G}
\label{eseven}
\ee
Thus the total energy is
\be
E\sim N_{\rm hole} E_{\rm hole}\sim {V\over R^2 G}
\label{elattice}
\ee
From this expression we have
\be
R\sim \left ( {V\over EG}\right )^\h
\ee
Substituting this in (\ref{entropylattice}) we find
\be
S\sim {1\over R} {V\over G}\sim \left ( {V\over EG}\right )^{-\h}{V\over G} \sim \sqrt{EV\over G}
\ee
Introducing a constant $K$ of order unity we get
\be \label{eq:EntropyFormul}
S=K\sqrt{EV\over G}
\ee
Noting that
\be
\rho={E\over V}
\ee
we see that
\be
S=K\sqrt{\rho\over G} \, V
\label{oneqq}
\ee

\b

This is a very interesting result; we make a few comments to amplify its nature:

\b

(a) The entropy $S$ in the box is {\it extensive}; i.e., it is proportional to $V$. Thus we can define an intensive entropy density
\be
s\equiv {S\over V}=K\sqrt{\rho\over G}
\label{density}
\ee
This relation is interesting, because of its contrast with the area law (\ref{sbek}). The area law had given rise to a general feeling that quantum gravity has `relatively few' states in a given region; this would be the case if the entropy was proportional to the surface area of the region rather than the volume. But we see that the area law is relevant only in situations where the spatial extent of the black hole microstates are not subject to any constraint; this would be the case for a black hole  in asymptotically flat space. In a cosmology matter fills all space uniformly, so a given microstate cannot expand freely to any size it wants: if it tries to expand it will run into the next microstate in a lattice like fig.\ref{fbh}(b). In this constrained situation we get an extensive entropy; i.e., $S$ is proportional to the volume rather than a surface area. 

\b

(b) One might wonder why the black holes in the lattice will not merge and produce a single larger hole. This cannot happen if we do not increase $V$, since we have already seen that a single hole in the volume $V$ has {\it less} entropy than the lattice of holes. In a cosmology $V$ can increase with time, and then smaller holes can indeed merge to larger holes.  It is natural to require that the expansion proceed at a rate governed by Einstein's equations with the equation of state (\ref{density}); under this expansion smaller holes will indeed merge to larger ones at a certain rate. Let us review the dynamics of this `black hole gas'.

  The first law of thermodynamics gives
  \be
  TdS=dE+pdV
  \ee
    Thus
  \be
  T=\left ( {\p S\over \p E}\right ) _V^{-1} = {2\over K} \sqrt{EG\over V}
  \ee
  \be
  p=T\left ( {\p S\over \p V }\right ) _E =  {E\over V}=\rho
  \label{eos}
  \ee
  Writing $p=w \rho$ we see that 
  \be
  w=1
  \label{weo}
  \ee
 Using matter with these properties in Einstein's equations, the cosmology (\ref{cosmology}) has the expansion 
\be
 a(t) = a_0 t^{1\over d}
\label{expansion}
\ee

This cosmology is termed the `black hole gas'. It was studied extensively in a series of papers by Banks and Fishler; see for example \cite{bf}. The same dynamics has also been argued for in many related ways. In \cite{veneziano} it was shown that  such an entropy density would be obtained for a closely packed gas of string states which are at the `Horowitz-Polchinski correspondence point' \cite{hp} (i.e., at the point where the string is large enough to be at the threshold of collapsing into a black hole). In \cite{fs} the relation (\ref{density}) was obtained by arguing that the entropy in a  cosmological horizon should be given by the area of that horizon.  In \cite{brusv} the notion of a causal connection scale was used to arrive at the same equation of state. The expression (\ref{one}) was obtained in \cite{sas1} by arguing for a `spacetime uncertainty relation'. In \cite{verlinde} a similar equation of state was argued to correspond to the Cardy formula for the density of states.

\subsection{The equation of state from duality invariance}

The black hole gas suggests a definite equation of state. Remarkably, we will find that this same equation of state arises when we proceed along a very different sounding line of thought.

Consider string theory, which lives in 9+1 spacetime dimensions. For concreteness, we take IIB string theory  (all different string theories can be mapped to each other by duality symmetries, so there is no loss of generality in doing this). As in the above discussion, we compactify the spatial direction to a torus with sides $L_1, \dots L_9$, so that
\be
V=\prod_{i=1}^9 L_i
\label{hone}
\ee
Consider again the limit of large energies $E$ in the box (\ref{hone}), and ask what is the expression for $S(E,V)$ predicted by string theory. This is a reasonably well defined question. It is true that Einstein's equations will force $V$ to increase according to the Hubble relation (\ref{otwo}). But in other situations in cosmology like the radiation phase of the standard model, or the string gas or brane gas phases in string theory, we have assumed that equilibrium is achieved more quickly than the expansion rate. This allows us to ask for $S(E,V)$. We will make the same assumption of being close to equilibrium, and ask for the equation of state predicted by string theory in the limit $E\r\infty$. We impose 3 requirements:

\b

 (i) String theory has a $T$ duality symmetry in each of the compact directions $x_i$. Under this symmetry,   the length $L_i$ changes to $1/L_i$ (as measured in the string frame), and the coupling $g$ changes in a certain way as well. Since $T$ duality is an exact symmetry of the theory, we require that the equation of state be invariant under this duality in each of the directions $x_i$.
 
 \b
 
 (ii) String theory has a $S$ duality symmetry where the string coupling $g$ goes to $1/g$. Again, since this is an exact symmetry, we require that the equation of state be invariant under this duality.
 
 \b
 
 (iii) In the limit where we reduce $E$ down to $E\sim E_{bh}$  (eq. (\ref{oone})), $S$ should reduce to $S\sim A/G$, the entropy of a single black hole in the box. This requirement will connect what we know about black hole entropy in asymptotically flat space to the equation of state we seek for high energy densities in a box. 
 
 \b
 
 Remarkably, there is a unique expression for $S$ that is consistent with these three requirements:
 \be \label{eq:EntropyFormulp}
S=K\sqrt{EV\over G}
\ee
with $K$ a constant of order unity. Thus we again recover the expression (\ref{eq:EntropyFormul}). This fact was observed in \cite{sas}. A detailed derivation can be found in \cite{masoumi1}.

 \subsection{Questions}
 
 The above considerations raise some basic questions for any fundamental approach to understanding the big bang:
 
 \b
 
 (i) Is there any way to {\it not} have the equation of state (\ref{density})? The argument of section \ref{secmaximum}  was based on looking for the most likely state based on entropic considerations. If we do not want a state determined by such a argument, then we need to invoke the instability of gravity that leads to a loss of equilibrium in situations like inflation. Doing this seems to be like asking for inflation at the planck scale rather than at the GUTS scale. If we do not have a mechanism to get inflation at the planck scale, then can we avoid having a hot big bang with the `black hole gas' evolution (\ref{expansion})? 
 
 \b
 
 (B) One might argue that inflation would wipe out any signals of a pre-inflation phase, so the question of dynamics before inflation is a somewhat academic one. But the initial conditions that lead to inflation must be determined by the state of the universe before inflation. In particular the question of how many dimensions should be compact, and what this compactification must be, are  questions that relate to the state before inflation. If we wish to get a probability distribution for different inflaton configurations, then do we have to obtain this from the state described by the equation of state (\ref{density})?
 
 \b
 
 (C) String theory lends further support to the equation of state (\ref{density}). The fact that duality invariance yields this relation is interesting. Further, string theory removes one of the barriers to thinking about the `back hole gas' as the state near the big bang. With the traditional picture of the hole it seemed that black hole microstates were something novel, different from the states of photons or strings that constitute a high density fluid.  But progress in black hole physics has taught us that black hole microstates are just like any other state in string theory. A vibrating string can be thought of as having its energy partitioned into two kinds of energy carriers: a  1-brane (the string) and momentum modes giving vibrations on this string. Each independent source of energy contributes $\sqrt{E}$ to the entropy, so we get the Hagedorn result 
 \be
 S\sim \sqrt{E} \sqrt{E}\sim E
 \ee
 With 3 kinds of charges (1-brane, 5-brane and momentum) we get black holes in 4+1 noncompact dimensions with entropy
 \be
 S\sim (\sqrt{E})^3=E^{3\over 2}
 \ee
 and with 4 kinds of charges we get black holes in 3+1 noncompact  dimensions having
 \be
 S\sim (\sqrt{E})^4=E^2
 \ee
 Thus there is no reason to allow the string gas as `normal' stringy matter and not also allow the gas of black holes that gives the equation of state (\ref{density}).

\section{Mapping the  information paradox to cosmology}\label{cosmopuzzle}

Consider a black hole with mass $M\gg m_p$. The horizon region is  a region of low curvature, with curvatures  of order
\be
{\mathcal R}\sim (GM)^{-2} \ll l_p^{-2}
\label{mone}
\ee
Thus one might expect that in the quantum theory one could use  semiclassical physics in the region around the horizon. That is, if we consider a ball of radius $r\lesssim M$ around the horizon then the state of the quantum fields $|\chi\rangle$ in this ball is to a first approximation the vacuum state
\be
\langle 0|\chi\rangle=1-O(\delta), ~~~\delta\ll 1
\label{four}
\ee
This assumption of the quantum state leads to the phenomenon of pair creation. One member of the pair (which we call $b$) escapes to infinity as Hawking radiation \cite{hawking}. The other member (called $c$) has net negative energy; it falls into the hole and lowers the mass $M$ to maintain energy conservation. The two members of the pair are in an entangled state. For our purposes we can model this entanglement by letting each quantum have two states $0$ and $1$, and letting the pair have the state
\be
|\psi\rangle_{pair}={1\over \sqrt{2}}\left (|0\rangle_b|0\rangle_c+|1\rangle_b|1\rangle_c\right ) +O(\epsilon)
\label{ltwo}
\ee
The leading order part of this state implies an entanglement $S_{ent}=\ln 2$ between the quanta $b$ and $c$.  Focusing on this leading part, we see that the entanglement between the radiation and the remaining hole keeps rising monotonically. This gives  a sharp puzzle near the endpoint of evaporation: how can there be a large entanglement of the radiation with a hole that has so little mass? This is the well known black hole information paradox.

One might think that this is a subtle problem involving very quantum states of matter, and some small correction to the state (\ref{ltwo}) arising from some unknown quantum gravity effects would remove the problematic entanglement. After all the number of pairs emitted by the hole is
\be
N_{pairs}\sim \left ( {M\over m_p}\right )^2\gg 1
\ee
and a small correction (denoted by the term $O(\epsilon)$ in (\ref{ltwo})) could cumulate over the large number of pairs to remove the entanglement between the radiation and the hole by the time we reach the end point of evaporation. But the small corrections theorem \cite{cern} says that this {\it cannot} happen. The entanglement after $N$ steps of emission must grow as
\be
 S_{ent}(N+1)>S_{ent}(N)+\ln 2-2\epsilon
 \label{lthree}
 \ee
 Thus the entanglement continues to rise monotonically; small corrections cannot make it turn around and head to zero. Put another way, we need {\it  order unity} corrections to semiclassical dynamics at the horizon to resolve the information paradox.
 
 This situation has led to two approaches to resolving the paradox:
 
 \b
 
 (i) {\it Fuzzballs:} In string theory it appears that bound states of strings and branes swell up to a size that depends on the number of quanta in the bound state, in such a way that the radius of the bound  state is always a little larger than horizon radius. These `fuzzball' states radiate from their surface like normal bodies (rather than by a pair creation process (\ref{ltwo})); this resolves the information paradox \cite{fuzzballs}.
 
 \b
 
 (ii) {\it Wormholes:} The horizon retains semiclassical physics in a suitable effective description of the quantum gravity variables. But there are nonlocal effects \cite{cool} (schematically denoted `wormhole' effects) which can be used to extract information nonlocally from the interior of the hole. 
 
 \b

Now let us turn to cosmology, and see if we can get a version of the information paradox in that setting. We will make certain plausible sounding assumptions, and see that these indeed lead us to a version of the information puzzle. We proceed in the following steps:

\b

\begin{figure}[H]
\begin{center}
 \includegraphics[scale=.8] {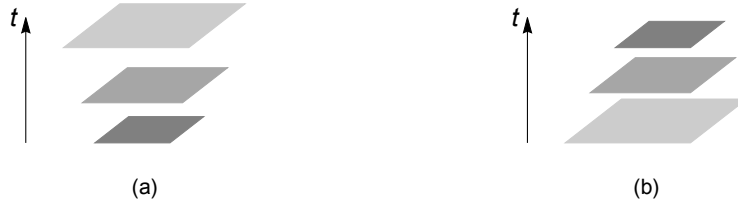}
\end{center}
\caption{(a) An expanding universe can be mapped by CPT to  (b) a collapsing universe. } 
\label{f2}
\end{figure}

(A)  Consider the flat dust cosmology
\be
ds^2=-dt^2 + a^2(t) [ dr^2+r^2 d\Omega_2^2]
\label{two}
\ee
where $a(t)=a_0 t^{2\over 3}$. We assume that {\it CPT invariance holds}. Then we can map this expanding universe (fig.\ref{f2}(a)) to a collapsing one (fig.\ref{f2}(b)).

\b

\begin{figure}[H]
\begin{center}
 \includegraphics[scale=1.1] {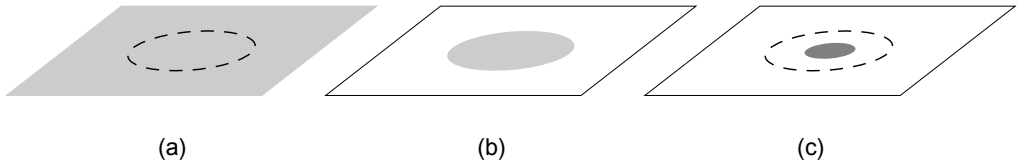}
\end{center}
\caption{ (a) A homogeneous collapsing flat cosmology (b) By the Birkoff theorem,  the dynamics of the marked ball cannot change if we remove the matter outside (c) Semiclassical dynamics suggests that this ball will pass through its horizon radius, but in actual fact the ball must tunnel to fuzzballs before this.} 
\label{f3}
\end{figure}

(B) Assume that {\it the Birkoff theorem  holds to  leading order even in the full quantum theory.} Consider a ball of ball of proper  radius $R(t)$ in the collapsing cosmology.   Then we can  replace the dust in the exterior  region $R>R(t)$ by  flat spacetime; by the Birkoff theorem this should not affect the dynamics of the ball in the  region $R<R(t)$   (fig.\ref{f3}(a,b)). Note  that in a dust cosmology there is no pressure across   the surface $R(t)$; this is important to allow the separation  of the outer and inner regions in this manner. 

\b

\b

(C) In the semiclassical picture of collapse, this  dust ball  will pass through  its horizon at some time $t$ to a radius satisfying    $R<2GM$. If we choose our initial ball large enough, then  this happens when the dust  density is still very low compared to  planck (fig.\ref{f3}(c)).

\b

\begin{figure}[h]
\begin{center}
 \includegraphics[scale=1] {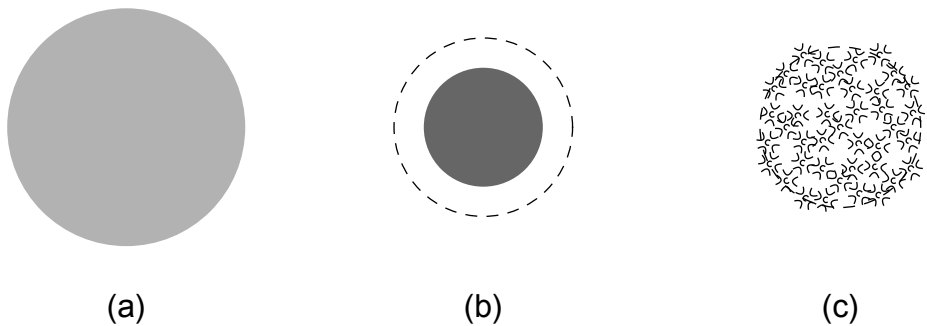}
\end{center}
\caption{ (a) A collapsing dust ball (b) The ball should never reach the semiclassically expected configuration (b), else we cannot solve the information puzzle (c) In string theory fuzzballs form instead when the ball reaches horizon radius. } 
\label{f1}
\end{figure}

With this, we have mapped the physics of a dust cosmology to the gravitational collapse of a finite dust ball. Let us recall what we need from the collapsing dust ball to resolve the information paradox. The fuzzball paradigm, for instance, says that a dust ball that is compressed to a size several times smaller than its horizon radius will relax, perhaps over a few crossing time scales, to horizon sized stringy objects called fuzzballs. We depict this transition in fig.\ref{f1}(c). But we do not have to focus on just the fuzzball approach here; {\it any} theory that seeks to modify dynamics when a horizon is formed will need to have novel violations of semiclassical dynamics when we reach the situation in fig.\ref{f1}(b).

Now consider the situation of the cosmology. We cannot see past our cosmological horizon today. But  in the past the horizon radius was smaller, and the dust denser. Thus looking back to earlier times we can find a ball shaped patch in the sky which has  the nature depicted in fig.\ref{f1}(b); i.e., the mass $M$ inside the ball of radius $R_b$ is such that the corresponding Schwarzschild radius $R_s$ (indicated by the dotted line in the figure) is much larger than the radius of the ball. Looking back to a time $T_1\approx 4\times 10^5 \, yrs$, say, when we had a dust cosmology, we find that we can see regions of the sky today where \cite{nagpur}
\be
{R_p\over R_s}\sim 2\times 10^{-4}\ll 1
\ee
But we do not see evidence of large departures from the semiclassical approximation in such dust balls. How should we reconcile this with any theory that seeks to modify semiclassical physics at the horizon scale to resolve the information paradox?

This seems to be a curious state of affairs, so let us locate the crux of the problem. Step (A) is standard; in fact it is often said that we live inside a white hole, which is just the time reverse of a black hole. Step (C)  just follows from the semiclassical evolution of the dust cosmology. The link between the two is step (B), where the Birkoff theorem is used to discard the outside of a collapsing region, and thus convert a collapsing cosmology to the  collapsing dust ball of the black hole problem. 

One might argue that the Birkoff theorem is classical, so there can be small quantum corrections to this theorem. Such corrections could allow the exterior of our dust ball to have some small quantum effects on the interior, and change cosmological evolution in the  region $R<R(\tau)$  slightly away from what we had in the black hole problem. But the small corrections theorem (\ref{lthree}) tells us that such small corrections cannot get us out of the information paradox. 

 What then could be the resolution of our conundrum? In the fuzzball paradigm, we have the vecro hypothesis, which says that virtual fluctuations of black holes exist for all radii $0<R_v<\infty$ in the Minkowski vacuum; the large action of such fluctuations for $R_v\gg l_p$ is offset by the very large entropy of  black holes states. (Here by `black hole states' here we mean configurations of the type that arise in the wavefunctionals of  fuzzballs.)
In an expanding cosmology, the situation is different from the situation in Minkowski space; the vecros have radii only in the range $0<R_v\lesssim H^{-1}$. This difference invalidates step (B) in the above discussion where we used the Birkoff theorem. The reason for this is that  just discarding the dust outside our ball does not convert this outer region to the Minkowski vacuum; we also need the vecro fluctuations with $R\gtrsim H^{-1}$ to be created before we can get this Minkowski spacetime. This resolves the above conflict, and allows the fuzzball resolution of the information paradox to be compatible with the absence of large violations of semiclassicality in the sky.

In general, the above discussion tells us that any proposal to resolve the information paradox should also agree with the constraints we get from observations of the cosmos.

\section{The cosmological constant}

It is often said that the natural value of the cosmological constant is
\be
\Lambda \sim l_p^{-4}
\label{tone}
\ee
where $l_p$ sets the UV cutoff on the modes of our quantum fields. But how do we arrive at this conclusion? 

The relation (\ref{tone}) seems to be indicated on dimensional grounds. We can also get such an expression if we add up the zero point energy of the harmonic oscillators that describe the field modes of, say, a scalar field, with a momentum cutoff $M\sim l_p^{-1}$. But as we will see note, the full stress tensor arising from such a cutoff is {\it not} of the form
\be
T_{\mu\nu}=-\Lambda \eta_{\mu\nu}
\label{tfourt}
\ee
We will follow the discussion of this issue in \cite{akhmedov, koksma, ossola}; a review can be found in \cite{martin}.

Consider for simplicity a massless scalar field in 1+1 spacetime dimensions
\be
L=-\h \p_\mu\phi \p^\mu\phi
\ee
We compactify the spatial direction to a circle of length $L$. 
The field can be expanded as
\be
\hat \phi= \sum_n {1\over \sqrt{2\omega_n}}{1\over \sqrt{L}}\left ( \hat a_n e^{ik_n x -i\omega_n t }+{\hat a_n} ^\dagger e^{-ik_n x +i\omega_n t}\right )
\label{ttwo}
\ee
where
\be
k_n={2\pi n\over L}, ~~n~{\rm integer} 
\ee
and
\be
\omega_n=|k_n|
\ee
The stress tensor is
\be
T_{\mu\nu}=\p_\mu \phi \p_\nu \phi -\h g_{\mu\nu} \p_\lambda\phi \p^\lambda \phi
\ee
Thus the energy density is
\be
\rho=T_{tt}=\h \dot\phi^2 +\h (\phi_{,x})^2
\ee
The pressure is also given by the same expression
\be
p=T_{xx}=\h \dot\phi^2 +\h (\phi_{,x})^2
\ee
This suggests that a naive computation will yield $\rho=p$, rather than the cosmological constant form form $\rho=-p$. Indeed, suppose we put a cutoff by letting
\be
0\le n \le n_{max}
\ee
for some large number $n_{max}$. Then using the field expression (\ref{ttwo}) we find
\be
\rho=\langle 0 |T_{tt} | 0 \rangle = {1\over 2L} \sum_{n=0}^{n_{max}} \omega_n \approx \h {2\pi\over L^2} {n_{max}^2\over 2}
\label{tthree}
\ee
and similarly
\be
p=\langle 0 |T_{xx} | 0 \rangle = {1\over 2L} \sum_{n=0}^{n_{max}} \omega_n \approx \h {2\pi\over L^2}{ n_{max}^2\over 2}
\label{tfour}
\ee
We can also get (\ref{tfour}) from (\ref{tthree}) by
\be
p=-{\p E\over \p L}= -{\p (\rho L)\over  \p  L}=-{\p\over \p L} \left (\h {2\pi\over L}{ n_{max}^2\over 2}\right ) =\h {2\pi\over L^2} { n_{max}^2\over 2}= \rho
\ee
To set the cutoff  to be order planck scale, we can take
\be
n_{max}={L\over l_p}
\label{tfive}
\ee
Then we will get
\be
\rho=p = \h {\pi\over l_p^2} 
\ee
This is the analogue of (\ref{tone}) on 1+1 spacetime dimensions. 

In 3+1 spacetime dimensions we get
\be
p={1\over 3} \rho
\ee
which again disagrees with the cosmological constant form $p=-\rho$. With a cutoff similar to (\ref{tfive}), we get
\be
p={1\over 3}\rho \sim {1\over l_p^4}
\ee
This agrees in its order of magnitude suggested by the expectation (\ref{tone}) that we had on dimensional grounds, but does not satisfy the Lorentz invariant form expected by Lorentz symmetry.  

Clearly the problem is that the cutoff breaks Lorentz invariance, and so the vacuum stress tensor is not coming out to be proportional to $\eta_{\mu\nu}$. One can remove the divergence by dimensional regularization, and then one indeed gets a Lorentz invariant result. For a massless field we get zero, while for a scalar field with mass $m$ we get (for 3+1 dimensions) \cite{akhmedov}
\be
\rho=-p=-{m^4\over 64\pi^2} \left ( \log ({  M^2\over m^2})+{3\over 2}\right )
\label{tten}
\ee
where $M$ is a UV cutoff. 

But dimensional regularization gives no physical picture of what excitations actually give rise to $\rho, p$. 
In particular it is unclear what kind of physics at the planck scale could lead to the sort of cutoff that would give the above expression obtained from dimensional regularization. The energy density in (\ref{tten}) is negative, and vanishes for $m=0$, which are not properties that agree with our initial intuition that the vacuum energy arises from the zero point energy of harmonic oscillators corresponding to the field modes.

The reason why the cutoff regularization failed to be Lorentz invariant is also clear. Consider the 1+1 dimensional example for simplicity. Suppose the circle had a length $L$. Placing a cutoff at the planck scale says that we should take
\be
n_{max}={L\over l_p}
\label{tel}
\ee
When we compute the pressure, we compute a partial derivative $p=-{\p E\over \p L}$. But when we increase $L$ by an amount $dL$, a larger number of planck lengths fit in the new length $L+dL$. If our basic principle is to maintain a cutoff at the planck scale $l_p$, then we must increase $n_{max}$ when we increase $L$. Thus instead of taking $n_{max}$ as a fixed number, we should let $n_{max}$ be given by the expression 
(\ref{tel}). Then from (\ref{tthree}) we have
\be
E=\rho L = \h {\pi L\over l_p^2} 
\ee
and
\be
p=-{\p E\over \p L}= -\h {\pi \over l_p^2} =-{E\over L}=-\rho
\ee
A similar analysis in 3+1 dimensions gives 
\be
p=-{\p E\over \p V}=-\rho
\ee
Thus we recover the desired Lorentz invariant form (\ref{tfourt}) of the stress energy tensor  for the vacuum.

But what is the physical meaning of setting $n_{max}$ to be determined by (\ref{tel})? When we increase $L$, we increase the number of lattice points on our circle, in order to keep the cutoff at length $l_p$. Since we have one field degree of freedom per lattice site, the number of variables defining our system are effectively changing when we increase $L$. 

The change in $L$ in computing the partial derivative ${\p E\over \p L}$ could be called a virtual change, but we will face the same issue in a more real way when we consider the expansion of the universe. As space expands, we will need more lattice sites to maintain a cutoff at length $l_p$, so the number of degrees of freedom defining the field will have to increase.

At this point we face a difficulty. The Hamiltonian formulation of quantum physics is set up to work on a Hilbert space of fixed dimension. Suppose our field theory could be modeled as a spin $\h$ degree of freedom at each lattice site. In a box of length $L$ along each axis $x,y,z$, we will have a Hilbert space of dimension
\be
{\cal N}=2^{({L\over l_p})^3}
\ee
A $L$ increases, ${\cal N}$ increases. How can such a system be described in a  Hamiltonian description?

If we have a continuous variable $\phi(x,y,z)$ at each lattice site, then the situation is more complicated. If we increase the number of lattice sites then we get new variables $\phi(x,y,z)$. Even in classical Hamiltonian mechanics, it is not clear how we would describe such dynamics. We could  discretize the allowed values of $\phi(x,y,z)$ at each lattice site  to get a Hilbert space that is finite dimensional. One could then imagine  a suitable reshuffle of these degrees of freedom when the universe expands so that the dimension ${\cal N}$ of the Hilbert space remains fixed. But as the universe expands to larger and larger values, there will come a point where there will be less that one state of the Hilbert space left at each lattice site. This situation would not describe a quantum field on a smooth manifold.\footnote{See \cite{expand} for a discussion of this issue.}

\b

In short, it is difficult to justify the naive expectation that a cutoff on vacuum field modes at the planck scale will give a vacuum stress-energy  $T_{\mu\nu}\sim \eta_{\mu\nu} l_p^{-4}$. There is no natural model of physics at the planck scale that generates such a stress energy. Formal manipulations like dimensional reduction give Lorentz covariant results, but the magnitude of $\Lambda$ is much smaller and there is no simple interpretation of the renormalization in terms of planck scale structure. 

Supersymmetry can cancel the cosmological constant, as happens for supersymmetric backgrounds in string theory. In that case on may not need to worry about how the field modes are cut off at high energy. But if supersymmetry is broken, as it appears to be on our world, then we need fine tuning again to set $\Lambda$ to be small. It is therefore not clear if supersymmetry provides a fundamental way to resolve the cosmological constant problem.

A different approach to the puzzle is suggested by the vecro picture \cite{vecro}. In this picture we are forced to consider virtual fluctuations of black hole microstates for all radii $R_v$ upto the cosmological horizon  $0<R_v\lesssim H^{-1}$. A higher vacuum energy leads to a smaller Hubble radius $H^{-1}$, and a correspondingly different vecro distribution. The vecro fluctuations of scale $R_v$ can be thought of correlations in the quantum gravity wavefunctional across distances $\sim R_v$; if we remove such correlations then the energy of the state rises. Thus one can imagine that the cosmological wavefunction gets driven to a configuration with larger $H^{-1}$, where more vecro correlations can arise and lower the overall energy. If such a resolution of the puzzle were correct, then the cosmological constant problem would not be a problem arising from planck scale structure but from the large scale structure of the gravitational vacuum. In this regard the situation would be similar to how we resolved puzzles with black holes. The vecro picture of the vacuum was needed  to resolve the information paradox and the `bags of gold' problem\footnote{See \cite{bags} for some statements and approaches regarding  this problem.} while maintaining causality; these resolutions act by changing the vacuum structure of the hole on the horizon scale. It would be satisfying if the same vacuum structure that is required to solve the black hole puzzle also resolves the cosmological constant puzzle and other questions in cosmology.

\section{Summary}

Cosmology provides perhaps the most interesting frontier of  physics today, where quantum theory, general relativity and observational results all converge. In this article we have noted that arguments based on very fundamental physical principles give rise to questions that any final theory of the universe will have to answer. Thus we should use these theoretical arguments as a complementary input to what we learn from observations of the sky.

\section*{Acknowledgements}

I would like to thank for useful discussions Robert Brandenberger,  Patrick Dasgupta,  Bin Guo, Anupam Mazumdar,  Emil Martinec, Ali Masoumi, Ken Olum,  Mohammed Sami,  Anjan Sen and Alex Vilenkin. This work is supported in part by DOE grant DE-SC0011726.


\end{document}